\UseRawInputEncoding
\documentclass[pdflatex,sn-mathphys]{sn-jnl}
\usepackage{graphicx}
\usepackage{amsmath}
\usepackage{amsfonts}
\usepackage{array}
\usepackage{siunitx}
\usepackage{bookmark}
\usepackage{natbib,hyperref}
\setcitestyle{round}
\usepackage{gensymb}
\usepackage{textgreek}
\usepackage{caption}
\jyear{2022}
\begin{document}

\newcommand{\red}{\textcolor{red}}

\title{\centering Measuring gravity with milligram levitated masses}

\author[1]{\fnm{Tim M.} \sur{Fuchs}}
\author[1]{\fnm{Dennis G.} \sur{Uitenbroek}}
\author[1]{\fnm{Jaimy} \sur{Plugge}}
\author[1]{\fnm{Noud} \sur{van Halteren}}
\author[1]{\fnm{Jean-Paul} \sur{van Soest}}
\author[3]{\fnm{Andrea} \sur{Vinante}} 
\author[2]{\fnm{Hendrik} \sur{Ulbricht}}
\author*[1]{\fnm{Tjerk H.} \sur{Oosterkamp}}\email{oosterkamp@physics.leidenuniv.nl}

\affil[1]{\orgdiv{Leiden Institute of Physics}, \orgname{Leiden University}, \orgaddress{P.O. Box 9504, 2300 RA Leiden, The Netherlands}}
\affil[2]{\orgdiv{School of Physics and Astronomy}, \orgname{University of Southampton}, \orgaddress{SO17 1BJ, Southampton, UK}}
\affil[3]{\orgdiv{Istituto di Fotonica e Nanotecnologie}, \orgname{CNR and Fondazione Bruno Kessler}, \orgaddress{I-38123 Povo, Trento, Italy}}

\abstract{Gravity differs from all other known fundamental forces since it is best described as a curvature of spacetime. For that reason it remains resistant to unifications with quantum theory. Gravitational interaction is fundamentally weak
and becomes prominent only at macroscopic scales. This means, we do
not know what happens to gravity in the microscopic regime where
quantum effects dominate, and whether quantum coherent effects of gravity become apparent. Levitated mechanical systems of mesoscopic size
offer a probe of gravity, while still allowing quantum control over their
motional state. This regime opens the possibility of table-top testing of
quantum superposition and entanglement in gravitating systems.
Here we show gravitational coupling between a levitated sub-millimeter
scale magnetic particle inside a type-I superconducting trap and kg
source masses, placed approximately half a meter away. Our results
extend gravity measurements to low gravitational forces of attonewton and underline the importance of levitated
mechanical sensors. }

\maketitle

\section{Introduction}\label{sec1}
    Einstein’s theory of general relativity (GR), our widely accepted theory of gravity, has seen different experimental confirmations~\cite{Einstein1916,walsh1979} by observing massive astronomical objects and their dynamics, most recently by the direct observation of gravitational waves from the merger of two black holes~\cite{abbott2016} and the imaging of a black hole by the event horizon telescope~\cite{akiyama2019}, as well as dedicated satellite missions for testing the basic principle of GR --- the equivalence principle~\cite{touboul2017} and frame dragging effects~\cite{everitt2011}. Laboratory experiments have been continuously increasing the sensitivity of gravity phenomena, including general relativistic effects in atom clocks and atom interferometers~\cite{bothwell2022,asenbaum2017}, tests of the equivalence principle~\cite{rosi2017,asenbaum2020}, precision measurements of Newton's constant~\cite{rosi2014,quinn2013} and tests of the validity of Newton's law at micrometre-scale distances~\cite{geraci2008,tan2020}. 
    
    However, gravity has never been tested for small masses and on the level of the Planck mass. Measurements of gravity from classical sources in laboratory table-top settings is contrasted by an increasing interest to study gravitational phenomena originating from quantum states of source masses, for example, in the form of the gravitational field generated by a quantum superposition state~\cite{bronstein2012republication,rickles2011role,bose2017,marletto2017,al2018optomechanical}. The effort ultimately aims at directly probing the interplay between quantum mechanics and general relativity in table-top experiments. Because quantum coherence is easily lost for increasing system size, it is important to isolate gravity as a coupling force for as small objects as possible, which in turn means to measure gravitational forces and interactions extremely precisely. 

    At the same time, massive quantum sensors are especially suited for tests in a regime with appreciable gravitational influences, which is favourable in probing fundamental decoherence mechanisms related to gravity \cite{R1,R2} or proposed  physical models of the wave function collapse~\cite{diosi2015,vinante2016,bassi2013} featuring the system mass explicitly, such as the continuous spontaneous localization (CSL) model~\cite{ghirardi1990} and the Di\'{o}si-Penrose model of gravitationally-induced collapse~\cite{diosi1987,penrose1996,oosterkamp2013}.

    An emerging technology for ultra-sensitive sensing is based on levitated mechanical systems. These can be used for the mechanical sensing of very weak forces and to probe quantum physics at increasing scales of mass (and space). In optical levitation schemes, the heating from trapping lasers is the most prominent source of noise. Worse, in any quantum experiment, they will provide a source of decoherence, greatly increasing the difficulty of creating macroscopic quantum states. In magnetically levitated systems, this pathway of decoherence is largely removed~\cite{Romero2021}.

    The extremely low damping of magnetic systems, combined with their relatively high mass and operation in low noise cryogenic environments, makes them well suited for mesoscopic probes of quantum mechanics and could provide a test to possible limits of the applicability of quantum mechanics to the macroscopic world~\cite{leggett2002,arndt2014}.
    Van Waarde et al.~\cite{Waarde2016} and Vinante et al.~\cite{vinante2020} have previously realised such magnetic levitation of sub-milligram particles, in which the motional state of the particle is read out by means of superconducting quantum interference device (SQUID) detection.

    In a recent publication, Westphal et al. have demonstrated gravitational coupling between two \SI{90}{mg}, \SI{1}{mm} radius, gold spheres, achieved off resonance at millihertz frequencies in a torsion balance-type geometry~\cite{aspelmeyer2021}. Recent work by Brack et al. ~\cite{brack2022} has shown the dynamical detection of gravitational coupling between two parallel beams of a  meter in size in the hertz regime. In this paper we present work with a \SI{2.4}{kg} source mass and a magnetically levitated sub-milligram test mass, giving a coupling of \SIrange{10}{30}{aN} with a force noise of $\SI{0.5}{fN/\sqrt{Hz}}$. This work provides an intermediate step towards an experiment where a small test mass senses the gravity sourced by a small source mass.

\section{Results}\label{sec:setup}
    The core of the setup is a type-I superconducting trap with a magnetic particle levitated therein, as shown in Fig.~\ref{fig1:setup}b.
    The trap is made of tantalum with a critical temperature of $T_c = \SI{4.48}{K}$. We perform the experiment at temperatures below \SI{100}{mK}.
    The trap has an elliptical shape (\SI{4.5}{mm}$~\times$~\SI{3.5}{mm}, with the height from bottom of the trap to the coil being \SI{4.7}{mm}) to confine the modes of the levitated magnetic particle to the axial system of the trap.
    The particle is composed of a set of three $\SI{0.25}{mm} \times \SI{0.25}{mm} \times \SI{0.25}{mm}$ Nd$_{2}$Fe$_{14}$B magnets that are magnetically attached North-to-South as also shown in Fig.~\ref{fig1:setup}b, and a spherical glass bead, \SI{0.25}{mm} radius, that is attached to the middle magnet using stycast. This bead is added to break the rotational symmetry of the zeppelin around the x-axis (angle $\gamma$).
    Typical remnant magnetisation of Nd$_{2}$Fe$_{14}$B magnets is in the order of \SI{1.4}{T}.
    The estimated mass of the full particle, as depicted in Fig.~\ref{fig1:setup}d, is \SI{0.43}{mg}.
    Using the infinite plane approximation of Vinante et al. \cite{vinante2020}, we calculate an expected z-mode frequency of \SI{27}{Hz} in this geometry.
    
    The motion of the particle results in a change of flux through a loop at the top of the trap (the pick-up loop), which is detected using a two-stage biased SQUID coupled inductively to the pick-up loop. The loop is positioned off-center so that the symmetry is broken, and all modes couple to the loop.
    A third loop positioned halfway between the SQUID input loop and the pick-up loop is coupled inductively to a calibration loop. This transformer is used to calibrate the energy coupling $\beta^2$ between the detection circuit and the degrees of motion of the zeppelin, providing calibrated motion of the zeppelin from the measured flux signal.
    This procedure is further described in Supplementary Materials C.
    
    The set-up is suspended from springs in a multi-stage mass spring system to shield the experiment from external vibrations, both vertical and lateral.
    The bottom three masses (one aluminum, two copper) are similar in weight to the experimental setup, with a lowest resonance frequency at \SI{0.9}{Hz}.
    Above that is a millikelvin mass spring system with a lowest resonance frequency of \SI{4.8}{Hz}.
    We refer to Ref.~\cite{wit2019} for more details on a near identical mass-spring system and its performance in a similar dilution refrigerator.
    This combination is suspended from the 1K plate by a long spring. 
    Thermalisation of the experiment is provided by a flattened silver wire, which is mechanically soft while providing a good thermal link.
    This entire system is depicted in Fig.~\ref{fig1:setup}A. 
    
    The cryostat as a whole is rigidly attached to a 25 metric ton concrete block, which is again placed on pneumatic dampers to limit vibrations coupling in from the building. The pulse tube cooler and the vacuum pumps for the circulation of the mixture are rigidly attached to the building through a second frame, and attached to the cryostat only by edge welded bellows and soft copper braiding to further limit external excitations from reaching the particle.
    
    To demonstrate the force sensitivity of the system and as a proof of concept for gravitational coupling in levitated magnetic systems, we utilized an electrically driven wheel with a set of three \SI{2.45}{kg} brass masses, placed equally spaced along the outer rim.
    This wheel was used to create a time dependent gravitational gradient at the resonant frequency of a selected mode of the zeppelin, in an effort to drive the motion gravitationally.
    The frequency of the masses was read out optically using a laser and photodiode, in which the masses act as a mechanical shutter.

    \begin{figure}[ht]
        \centering
        \includegraphics[width=0.8\textwidth]{Setup/SetupSchematic.pdf}
        \caption{\textbf{Schematic depiction of the experimental setup.}\\
        \textbf{\emph{A}}: Multi-stage mass spring system to isolate from external vibrations, as discussed in the text. Electromagnetic shielding of the trap is discussed in the Supplementary Material A.
        \textbf{\emph{B}}: Conventions for degrees of freedom adopted from Vinante et al.~\cite{vinante2020}. Detection by SQUID as discussed in the text.  Calibration loop as discussed in the text and Supp.Mat. C. 
        \textbf{\emph{C}}: An image of: the dilution refrigerator used for the experiments, including the multi-stage mass spring system. 
        \textbf{\emph{D}}: The magnetic particle, composed of three $\SI{0.25}{mm} \times \SI{0.25}{mm} \times \SI{0.25}{mm}$ Nd$_{2}$Fe$_{14}$B magnets (SuperMagnetMan, C0005-10) magnetically attached end-to-end and a single spherical glass bead with a \SI{0.25}{mm} radius attached using Stycast to the middle of the magnets, which is used to break the symmetry of the $\gamma$ mode.
        \textbf{\emph{E}}: The trap, as placed in the aluminium holder without the shielding cylinder. The aluminium foil envelope provides additional electro-magnetic shielding between the calibration transformer and the pick-up loop.
        Further details and images of the setup are shown in the Supplementary Material A.
        }\label{fig1:setup}
    \end{figure}

\begin{figure}[ht]%
        \centering
        \includegraphics[width=\textwidth]{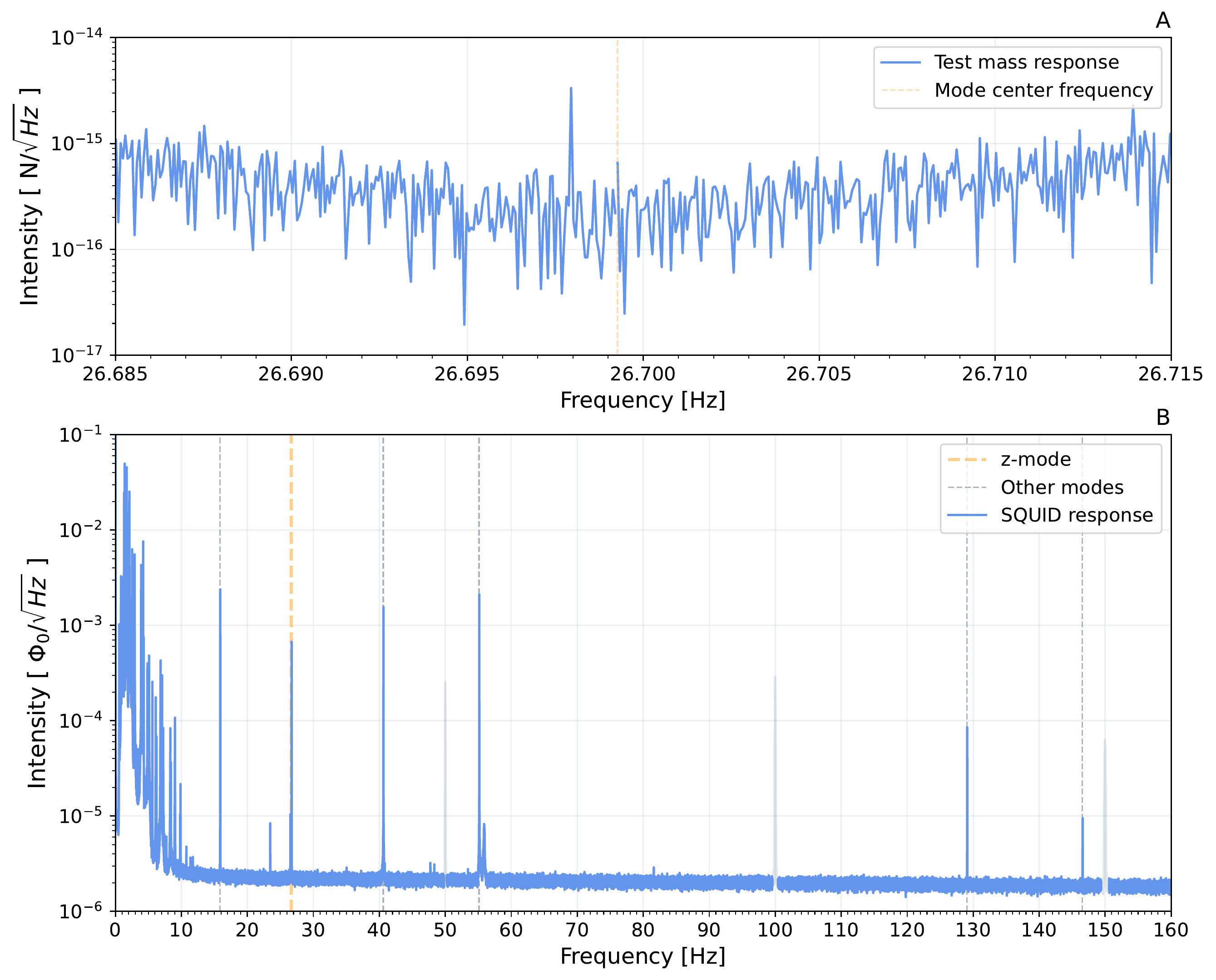}
        \caption{\textbf{Levitated Particle Spectra.}\\
        \textbf{\emph{A}}: Force noise of the \SI{27}{Hz} mode when gravitationally driven at \SI{1.3}{mHz} detuning, overnight. \textbf{\emph{B}}: Typical resonator power spectrum. In this figure we have greyed out the regular \SI{50}{Hz} European electrical noise. This noise typically has a similar power to the particle resonances.}\label{fig2:spectrum}
    \end{figure}

    In figure~\ref{fig2:spectrum}B. we show the uncalibrated spectrum.
    We clearly observe the six different modes corresponding to the three translational modes, and three rotational modes respectively. 
    Furthermore, we see a distinct peak at 27 Hz, which we attribute to the z-mode as discussed in the beginning of this section.
    These modes were validated and calibrated by performing a magnetic drive, excited by a flux injected through the calibration transformer, as shown in figure~\ref{fig1:setup}B. 
    
    Using this magnetic drive, we determine the decay time of the modes during the subsequent ringdown. For the \SI{26.7}{Hz} mode, we find a lower bound  $\tau = \SI{1.09e5}{s}$, or a Q factor of $Q = \SI{9.13e6}{}$. This procedure is further discussed in the Supplementary Material B. For the other modes, we arrive at Q factors that are about an order of magnitude lower.
    
    We test the force sensitivity of our mode by driving the \SI{26.7}{Hz} mode using the brass masses of the mass-wheel. 
    The resulting excitation at one position of the wheel in the force spectrum, is shown in figure~\ref{fig2:spectrum}A. 
    The calibration of this spectrum is discussed in the supplementary materials. 
    The resulting force noise of this z-mode is approximately $\SI{0.5}{fN/\sqrt{Hz}}$,  or equivalently, a displacement noise of $\SI{60}{pm/\sqrt{Hz}}$, in an \SI{8}{mHz} bandwidth centered around the orange dotted line that indicates the frequency of the resonance. 
    Equivalently, we can determine the motion of the trap in which the particle is levitated by dividing the force noise by the spring constant of the confinement potential that keeps the particle around its equilibrium height. 
    The spring constant for the z-mode was determined to be k = \SI{12e{-3}}{N/m}, resulting in a trap displacement noise of $\SI{30}{fm/\sqrt{Hz}}$.
    This vibrational noise is not yet thermally limited, but rather corresponds to a mode temperature of 3 K, which we attribute to the limits of the vibration isolation inside the cryostat.
    
\begin{figure}[ht]%
        \centering
        \includegraphics[width=\textwidth]{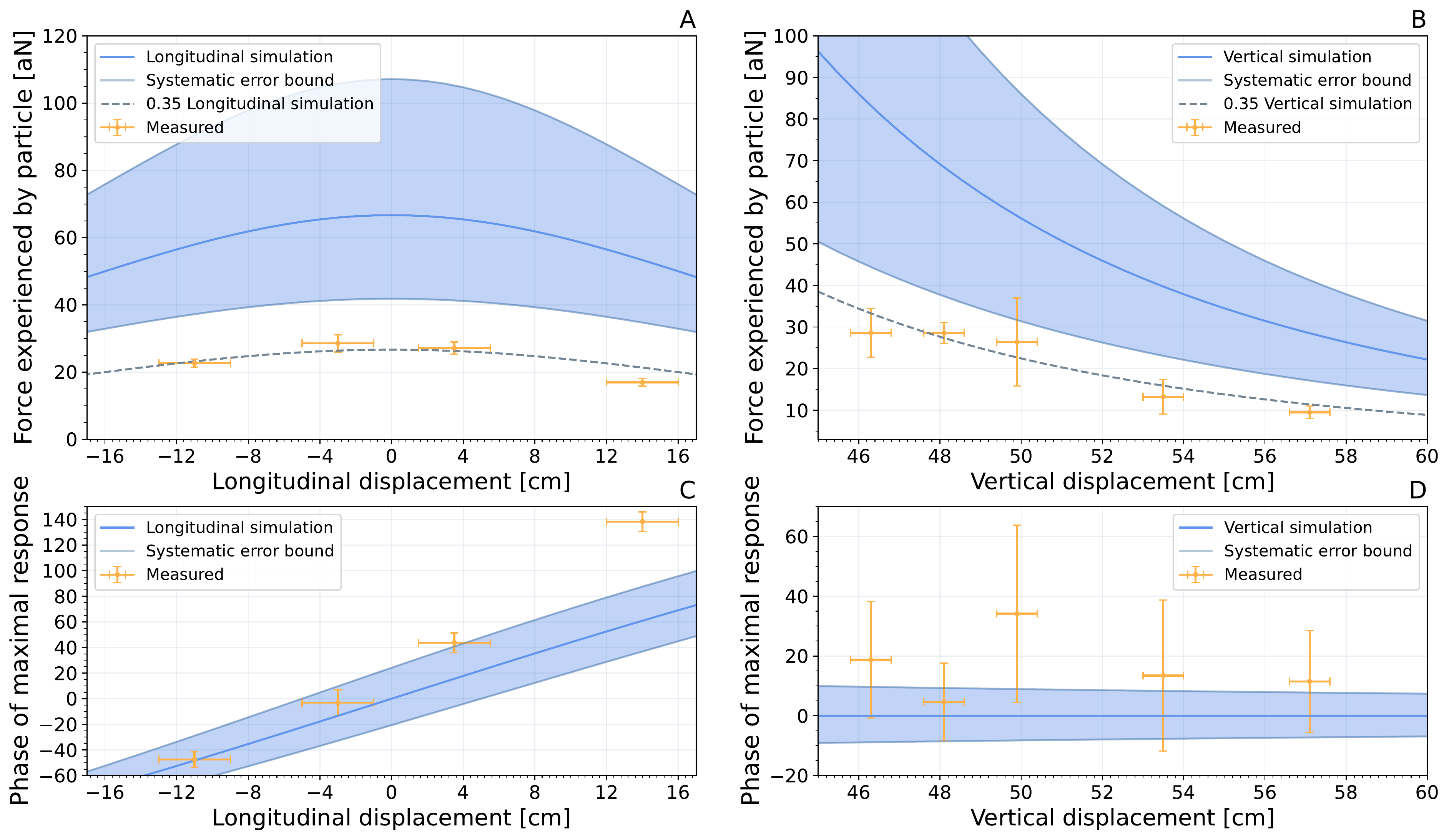}
        \caption{\textbf{Response to gravitational drive as function of separation.}\\
        \textbf{\emph{A}}: Force experienced by the mechanical resonator as a result from drive using the mass wheel, for different lateral displacement of the mass wheel relative to the particle. The dashed blue line represents the simulated gravitational force at the position of the magnetic particle as discussed in the main text. The blue area denotes the systematic uncertainty discussed in the text. A second, dashed line is plotted, which has a scaling factor of 0.35 applied, that seems to agree with the data more closely, as discussed in the main text. \textbf{\emph{B}}: Similar to A, but now for a vertical displacement of the mass wheel relative to the particle, when the wheel is kept centered below the particle. Systematic bounds as discussed in the text.   \textbf{\emph{C}}: Here we see the wheel phase at which the magnetic particle experiences the strongest force, plotted against longitudinal displacement. \textbf{\emph{D}}: Similar to C, but now for a vertical displacement.\\}\label{fig3:force}
    \end{figure}

    In figure~\ref{fig3:force} we show the measured gravitational interaction for different displacements of the wheel, using the method described in the Supplementary Material E. We also plot the phase of the masses along the wheels rotation for which the particle experiences maximal force, see figure S2. For the longitudinal displacement, the vertical displacement was held at \SI{48}{} $\pm$ \SI{4}{} centimeter. In the run of vertical positions, the wheel was kept centered with respect to the trap. Included is the expected gravitational signal at the location of the magnetic particle for the z-mode of the particle, which was calculated from an analytical simulation where the mass was taken to consist of multiple point masses. From this same simulation, a systematic error bound was derived, based on an estimated systematic error, for the longitudinal run, of $\pm$5 centimeter longitudinal, $\pm$3 centimeter lateral, and $\pm$4 centimeter vertical, which were estimated from the geometry of the wheel, the mass spring system and the systems used to measure the displacements. For the vertical run, the bounds are $\pm$2 centimeter in each principal direction, based on the increased stability of the system under vertical displacement.

    The observed signal agrees with the simulated force signal to within a factor 0.35 with a standard error of 0.02, which was determined by means of an orthogonal distance regression fit to the data.

\section{Discussion}\label{sec:conclusions}    
    We attribute this constant factor to the effect of the wheel on the motion of the trap and its holder. The trap and its platform also experience a gravitational pull.  For the sake of its vibration isolation, the trap and its platform is suspended by springs and thus the platform is set in motion by the gravitational pull on it. The amplitude of this motion is extra small, however, because the frequency of the gravitational drive is approximately a factor of 10 above the resonance of the suspension. This naturally gives rise to a 180\textdegree\ phase shift in the response of the trap's motion. The small motion of the trap cause the walls of the trap to exert a small force on the particle. Because of the 180 degree phase shift, this force is out of phase with the gravitational pull that the particle experiences. Thus this leads to a suppression of the particle response. The suppression depends on the total force on the platform, the gravitational gradient due to the wheel, the angle which the platform makes with respect to the horizontal etc. Additional complicating factors are the changing stiffness of the SQUID cable when cooling down, which can lead to a significant tilt of the platform.

    We demonstrate the detection of a \SI{30}{aN} gravitational signal at \SI{27}{Hz} and a damping linewidth as low as $\gamma/2\pi = \SI{2.9}{\mu Hz}$, with a \SI{0.43}{mg} test mass, paving the way for future experiments in which both source and test mass are in this regime. This work could be used to derive a more stringent bound on dissipative collapse models. Furthermore it provides a promising platform to test for possible deviations from inverse square force laws and fifth force models~\cite{Blakemore2021, smullin2005}, theories of modified Newtonian dynamics~\cite{milgrom1983, bekenstein2004} and other extensions of the standard model~\cite{carney2021}.
    
    By ensuring that the pick-up loop is placed off-centre with respect to the trap and by breaking the rotational symmetry in \textgamma\ we demonstrate detection of all six mechanical modes, in comparison to earlier work. As we will discuss in future work, this is critical to the stability of the mode under test due to non-linear mixing between the different modes.
    
    With a mode temperature of 3 kelvin compared to an operating temperature of 30 millikelvin, we are currently not yet thermally limited.
    It would require another 20 dB of vibration isolation to reach thermal motion.
    
    By using a second particle in a different trap as source mass, or a similar construction, this work paves the way towards easily scalable measurements of gravitational coupling in the hertz regime and with source masses at Planck mass level, ultimately allowing for testing gravity in a yet unexplored low-mass regime and pushing into the quantum controlled domain. Coupling of the detection SQUIDs in this scheme to an superconducting LC circuit would provide a means of inserting single microwave photons, providing access to the toolbox of quantum state manipulation. This would further extend this work towards truly macroscopic superposition measurements and possibly gravitationally-induced entanglement.

\backmatter

%

\bmhead{Acknowledgments}
We thank K. Heeck, B. Hensen, P. Numberi, G. Koning, C. Timberlake, E. Simcox and M. Camp for useful discussions and experimental help.

\bmhead{Funding} 
This work was supported by the NWO grant OCENW.GROOT.2019.088., the QuantERA grant LEMAQUME, funded by the QuantERA
II ERA-NET Cofund in Quantum Technologies implemented within the European Union’s Horizon
2020 Programme. Further, we would like to thank for support the UK funding agency EPSRC
under grants EP/W007444/1, EP/V035975/1 and EP/V000624/1, the Leverhulme Trust (RPG-
2022-57), the EU Horizon 2020 FET-Open project TeQ (766900) and the EU Horizon Europe EIC Pathfinder project
QuCoM (10032223).

\bmhead{Author Contributions}
TMF, DGU, JP, NvH, JPvS, AV, HU and THO developed the experimental hardware and procedure and contributed to the manuscript; TMF, DGU, JP, NvH and THO collected the data and contributed to data analysis.

\bmhead{Competing Interests}
All authors declare that they have no competing interests.

\bmhead{Data Availability}
All data needed to evaluate the conclusions in the paper are present in the paper and/or the Supplementary Materials. Raw data (timetraces of lock-in output) is available on Zenodo: \url{https://zenodo.org/records/10300430}


\clearpage

\bibliography{main}


\begin{thebibliography}{41}
\ifx \bisbn   \undefined \def \bisbn  #1{ISBN #1}\fi
\ifx \binits  \undefined \def \binits#1{#1}\fi
\ifx \bauthor  \undefined \def \bauthor#1{#1}\fi
\ifx \batitle  \undefined \def \batitle#1{#1}\fi
\ifx \bjtitle  \undefined \def \bjtitle#1{#1}\fi
\ifx \bvolume  \undefined \def \bvolume#1{\textbf{#1}}\fi
\ifx \byear  \undefined \def \byear#1{#1}\fi
\ifx \bissue  \undefined \def \bissue#1{#1}\fi
\ifx \bfpage  \undefined \def \bfpage#1{#1}\fi
\ifx \blpage  \undefined \def \blpage #1{#1}\fi
\ifx \burl  \undefined \def \burl#1{\textsf{#1}}\fi
\ifx \doiurl  \undefined \def \doiurl#1{\url{https://doi.org/#1}}\fi
\ifx \betal  \undefined \def \betal{\textit{et al.}}\fi
\ifx \binstitute  \undefined \def \binstitute#1{#1}\fi
\ifx \binstitutionaled  \undefined \def \binstitutionaled#1{#1}\fi
\ifx \bctitle  \undefined \def \bctitle#1{#1}\fi
\ifx \beditor  \undefined \def \beditor#1{#1}\fi
\ifx \bpublisher  \undefined \def \bpublisher#1{#1}\fi
\ifx \bbtitle  \undefined \def \bbtitle#1{#1}\fi
\ifx \bedition  \undefined \def \bedition#1{#1}\fi
\ifx \bseriesno  \undefined \def \bseriesno#1{#1}\fi
\ifx \blocation  \undefined \def \blocation#1{#1}\fi
\ifx \bsertitle  \undefined \def \bsertitle#1{#1}\fi
\ifx \bsnm \undefined \def \bsnm#1{#1}\fi
\ifx \bsuffix \undefined \def \bsuffix#1{#1}\fi
\ifx \bparticle \undefined \def \bparticle#1{#1}\fi
\ifx \barticle \undefined \def \barticle#1{#1}\fi
\bibcommenthead
\ifx \bconfdate \undefined \def \bconfdate #1{#1}\fi
\ifx \botherref \undefined \def \botherref #1{#1}\fi
\ifx \url \undefined \def \url#1{\textsf{#1}}\fi
\ifx \bchapter \undefined \def \bchapter#1{#1}\fi
\ifx \bbook \undefined \def \bbook#1{#1}\fi
\ifx \bcomment \undefined \def \bcomment#1{#1}\fi
\ifx \oauthor \undefined \def \oauthor#1{#1}\fi
\ifx \citeauthoryear \undefined \def \citeauthoryear#1{#1}\fi
\ifx \endbibitem  \undefined \def \endbibitem {}\fi
\ifx \bconflocation  \undefined \def \bconflocation#1{#1}\fi
\ifx \arxivurl  \undefined \def \arxivurl#1{\textsf{#1}}\fi
\csname PreBibitemsHook\endcsname

\bibitem{Einstein1916}
\begin{barticle}
\bauthor{\bsnm{{Einstein}}, \binits{A.}}:
\batitle{{Die Grundlage der allgemeinen Relativit{\"a}tstheorie}}.
\bjtitle{Annalen der Physik}
\bvolume{354}(\bissue{7}),
\bfpage{769}--\blpage{822}
(\byear{1916}).
\doiurl{10.1002/andp.19163540702}
\end{barticle}
\endbibitem

\bibitem{walsh1979}
\begin{barticle}
\bauthor{\bsnm{{Walsh}}, \binits{D.}},
\bauthor{\bsnm{{Carswell}}, \binits{R.F.}},
\bauthor{\bsnm{{Weymann}}, \binits{R.J.}}:
\batitle{{0957+561 A, B: twin quasistellar objects or gravitational lens?}}
\bjtitle{nature}
\bvolume{279},
\bfpage{381}--\blpage{384}
(\byear{1979}).
\doiurl{10.1038/279381a0}
\end{barticle}
\endbibitem

\bibitem{abbott2016}
\begin{barticle}
\bauthor{\bsnm{Abbott}, \binits{B.P.}},
\bauthor{\bsnm{Abbott}, \binits{R.}},
\bauthor{\bsnm{Abbott}, \binits{T.}},
\bauthor{\bsnm{Abernathy}, \binits{M.}},
\bauthor{\bsnm{Acernese}, \binits{F.}},
\bauthor{\bsnm{Ackley}, \binits{K.}},
\bauthor{\bsnm{Adams}, \binits{C.}},
\bauthor{\bsnm{Adams}, \binits{T.}},
\bauthor{\bsnm{Addesso}, \binits{P.}},
\bauthor{\bsnm{Adhikari}, \binits{R.}}, \betal:
\batitle{Observation of gravitational waves from a binary black hole merger}.
\bjtitle{Physical review letters}
\bvolume{116}(\bissue{6}),
\bfpage{061102}
(\byear{2016})
\end{barticle}
\endbibitem

\bibitem{akiyama2019}
\begin{barticle}
\bauthor{\bsnm{Akiyama}, \binits{K.}},
\bauthor{\bsnm{Alberdi}, \binits{A.}},
\bauthor{\bsnm{Alef}, \binits{W.}},
\bauthor{\bsnm{Asada}, \binits{K.}},
\bauthor{\bsnm{Azulay}, \binits{R.}},
\bauthor{\bsnm{Baczko}, \binits{A.-K.}},
\bauthor{\bsnm{Ball}, \binits{D.}},
\bauthor{\bsnm{Balokovi{\'c}}, \binits{M.}},
\bauthor{\bsnm{Barrett}, \binits{J.}},
\bauthor{\bsnm{Bintley}, \binits{D.}}, \betal:
\batitle{First m87 event horizon telescope results. iv. imaging the central supermassive black hole}.
\bjtitle{The Astrophysical Journal Letters}
\bvolume{875}(\bissue{1}),
\bfpage{4}
(\byear{2019})
\end{barticle}
\endbibitem

\bibitem{touboul2017}
\begin{barticle}
\bauthor{\bsnm{Touboul}, \binits{P.}},
\bauthor{\bsnm{M{\'e}tris}, \binits{G.}},
\bauthor{\bsnm{Rodrigues}, \binits{M.}},
\bauthor{\bsnm{Andr{\'e}}, \binits{Y.}},
\bauthor{\bsnm{Baghi}, \binits{Q.}},
\bauthor{\bsnm{Berg{\'e}}, \binits{J.}},
\bauthor{\bsnm{Boulanger}, \binits{D.}},
\bauthor{\bsnm{Bremer}, \binits{S.}},
\bauthor{\bsnm{Carle}, \binits{P.}},
\bauthor{\bsnm{Chhun}, \binits{R.}}, \betal:
\batitle{Microscope mission: first results of a space test of the equivalence principle}.
\bjtitle{Physical review letters}
\bvolume{119}(\bissue{23}),
\bfpage{231101}
(\byear{2017})
\end{barticle}
\endbibitem

\bibitem{everitt2011}
\begin{barticle}
\bauthor{\bsnm{Everitt}, \binits{C.F.}},
\bauthor{\bsnm{DeBra}, \binits{D.}},
\bauthor{\bsnm{Parkinson}, \binits{B.}},
\bauthor{\bsnm{Turneaure}, \binits{J.}},
\bauthor{\bsnm{Conklin}, \binits{J.}},
\bauthor{\bsnm{Heifetz}, \binits{M.}},
\bauthor{\bsnm{Keiser}, \binits{G.}},
\bauthor{\bsnm{Silbergleit}, \binits{A.}},
\bauthor{\bsnm{Holmes}, \binits{T.}},
\bauthor{\bsnm{Kolodziejczak}, \binits{J.}}, \betal:
\batitle{Gravity probe b: final results of a space experiment to test general relativity}.
\bjtitle{Physical Review Letters}
\bvolume{106}(\bissue{22}),
\bfpage{221101}
(\byear{2011})
\end{barticle}
\endbibitem

\bibitem{bothwell2022}
\begin{barticle}
\bauthor{\bsnm{Bothwell}, \binits{T.}},
\bauthor{\bsnm{Kennedy}, \binits{C.J.}},
\bauthor{\bsnm{Aeppli}, \binits{A.}},
\bauthor{\bsnm{Kedar}, \binits{D.}},
\bauthor{\bsnm{Robinson}, \binits{J.M.}},
\bauthor{\bsnm{Oelker}, \binits{E.}},
\bauthor{\bsnm{Staron}, \binits{A.}},
\bauthor{\bsnm{Ye}, \binits{J.}}:
\batitle{Resolving the gravitational redshift across a millimetre-scale atomic sample}.
\bjtitle{Nature}
\bvolume{602}(\bissue{7897}),
\bfpage{420}--\blpage{424}
(\byear{2022})
\end{barticle}
\endbibitem

\bibitem{asenbaum2017}
\begin{barticle}
\bauthor{\bsnm{Asenbaum}, \binits{P.}},
\bauthor{\bsnm{Overstreet}, \binits{C.}},
\bauthor{\bsnm{Kovachy}, \binits{T.}},
\bauthor{\bsnm{Brown}, \binits{D.D.}},
\bauthor{\bsnm{Hogan}, \binits{J.M.}},
\bauthor{\bsnm{Kasevich}, \binits{M.A.}}:
\batitle{Phase shift in an atom interferometer due to spacetime curvature across its wave function}.
\bjtitle{Physical review letters}
\bvolume{118}(\bissue{18}),
\bfpage{183602}
(\byear{2017})
\end{barticle}
\endbibitem

\bibitem{rosi2017}
\begin{barticle}
\bauthor{\bsnm{Rosi}, \binits{G.}},
\bauthor{\bsnm{D'Amico}, \binits{G.}},
\bauthor{\bsnm{Cacciapuoti}, \binits{L.}},
\bauthor{\bsnm{Sorrentino}, \binits{F.}},
\bauthor{\bsnm{Prevedelli}, \binits{M.}},
\bauthor{\bsnm{Zych}, \binits{M.}},
\bauthor{\bsnm{Brukner}, \binits{{\v{C}}.}},
\bauthor{\bsnm{Tino}, \binits{G.}}:
\batitle{Quantum test of the equivalence principle for atoms in coherent superposition of internal energy states}.
\bjtitle{Nature communications}
\bvolume{8}(\bissue{1}),
\bfpage{15529}
(\byear{2017})
\end{barticle}
\endbibitem

\bibitem{asenbaum2020}
\begin{barticle}
\bauthor{\bsnm{Asenbaum}, \binits{P.}},
\bauthor{\bsnm{Overstreet}, \binits{C.}},
\bauthor{\bsnm{Kim}, \binits{M.}},
\bauthor{\bsnm{Curti}, \binits{J.}},
\bauthor{\bsnm{Kasevich}, \binits{M.A.}}:
\batitle{Atom-interferometric test of the equivalence principle at the 10- 12 level}.
\bjtitle{Physical Review Letters}
\bvolume{125}(\bissue{19}),
\bfpage{191101}
(\byear{2020})
\end{barticle}
\endbibitem

\bibitem{rosi2014}
\begin{barticle}
\bauthor{\bsnm{Rosi}, \binits{G.}},
\bauthor{\bsnm{Sorrentino}, \binits{F.}},
\bauthor{\bsnm{Cacciapuoti}, \binits{L.}},
\bauthor{\bsnm{Prevedelli}, \binits{M.}},
\bauthor{\bsnm{Tino}, \binits{G.}}:
\batitle{Precision measurement of the newtonian gravitational constant using cold atoms}.
\bjtitle{Nature}
\bvolume{510}(\bissue{7506}),
\bfpage{518}--\blpage{521}
(\byear{2014})
\end{barticle}
\endbibitem

\bibitem{quinn2013}
\begin{barticle}
\bauthor{\bsnm{Quinn}, \binits{T.}},
\bauthor{\bsnm{Parks}, \binits{H.}},
\bauthor{\bsnm{Speake}, \binits{C.}},
\bauthor{\bsnm{Davis}, \binits{R.}}:
\batitle{Improved determination of g using two methods}.
\bjtitle{Physical Review Letters}
\bvolume{111}(\bissue{10}),
\bfpage{101102}
(\byear{2013})
\end{barticle}
\endbibitem

\bibitem{geraci2008}
\begin{barticle}
\bauthor{\bsnm{Geraci}, \binits{A.A.}},
\bauthor{\bsnm{Smullin}, \binits{S.J.}},
\bauthor{\bsnm{Weld}, \binits{D.M.}},
\bauthor{\bsnm{Chiaverini}, \binits{J.}},
\bauthor{\bsnm{Kapitulnik}, \binits{A.}}:
\batitle{Improved constraints on non-newtonian forces at 10 microns}.
\bjtitle{Physical Review D}
\bvolume{78}(\bissue{2}),
\bfpage{022002}
(\byear{2008})
\end{barticle}
\endbibitem

\bibitem{tan2020}
\begin{barticle}
\bauthor{\bsnm{Tan}, \binits{W.-H.}},
\bauthor{\bsnm{Du}, \binits{A.-B.}},
\bauthor{\bsnm{Dong}, \binits{W.-C.}},
\bauthor{\bsnm{Yang}, \binits{S.-Q.}},
\bauthor{\bsnm{Shao}, \binits{C.-G.}},
\bauthor{\bsnm{Guan}, \binits{S.-G.}},
\bauthor{\bsnm{Wang}, \binits{Q.-L.}},
\bauthor{\bsnm{Zhan}, \binits{B.-F.}},
\bauthor{\bsnm{Luo}, \binits{P.-S.}},
\bauthor{\bsnm{Tu}, \binits{L.-C.}}, \betal:
\batitle{Improvement for testing the gravitational inverse-square law at the submillimeter range}.
\bjtitle{Physical Review Letters}
\bvolume{124}(\bissue{5}),
\bfpage{051301}
(\byear{2020})
\end{barticle}
\endbibitem

\bibitem{bronstein2012republication}
\begin{barticle}
\bauthor{\bsnm{Bronstein}, \binits{M.}}:
\batitle{Republication of: Quantum theory of weak gravitational fields}.
\bjtitle{General Relativity and Gravitation}
\bvolume{44}(\bissue{1}),
\bfpage{267}--\blpage{283}
(\byear{2012})
\end{barticle}
\endbibitem

\bibitem{rickles2011role}
\begin{bchapter}
\bauthor{\bsnm{Rickles}, \binits{D.}},
\bauthor{\bsnm{DeWitt}, \binits{C.M.}}:
\bctitle{The role of gravitation in physics: report from the 1957 chapel hill conference}.
In: \bbtitle{The Role of Gravitation in Physics: Report from the 1957 Chapel Hill Conference}
(\byear{2011})
\end{bchapter}
\endbibitem

\bibitem{bose2017}
\begin{barticle}
\bauthor{\bsnm{Bose}, \binits{S.}},
\bauthor{\bsnm{Mazumdar}, \binits{A.}},
\bauthor{\bsnm{Morley}, \binits{G.W.}},
\bauthor{\bsnm{Ulbricht}, \binits{H.}},
\bauthor{\bsnm{Toro{\v{s}}}, \binits{M.}},
\bauthor{\bsnm{Paternostro}, \binits{M.}},
\bauthor{\bsnm{Geraci}, \binits{A.A.}},
\bauthor{\bsnm{Barker}, \binits{P.F.}},
\bauthor{\bsnm{Kim}, \binits{M.}},
\bauthor{\bsnm{Milburn}, \binits{G.}}:
\batitle{Spin entanglement witness for quantum gravity}.
\bjtitle{Physical review letters}
\bvolume{119}(\bissue{24}),
\bfpage{240401}
(\byear{2017})
\end{barticle}
\endbibitem

\bibitem{marletto2017}
\begin{barticle}
\bauthor{\bsnm{Marletto}, \binits{C.}},
\bauthor{\bsnm{Vedral}, \binits{V.}}:
\batitle{Gravitationally induced entanglement between two massive particles is sufficient evidence of quantum effects in gravity}.
\bjtitle{Physical review letters}
\bvolume{119}(\bissue{24}),
\bfpage{240402}
(\byear{2017})
\end{barticle}
\endbibitem

\bibitem{al2018optomechanical}
\begin{barticle}
\bauthor{\bsnm{Al~Balushi}, \binits{A.}},
\bauthor{\bsnm{Cong}, \binits{W.}},
\bauthor{\bsnm{Mann}, \binits{R.B.}}:
\batitle{Optomechanical quantum cavendish experiment}.
\bjtitle{Physical Review A}
\bvolume{98}(\bissue{4}),
\bfpage{043811}
(\byear{2018})
\end{barticle}
\endbibitem

\bibitem{R1}
\begin{barticle}
\bauthor{\bsnm{{Pikovski}}, \binits{I.}},
\bauthor{\bsnm{{Zych}}, \binits{M.}},
\bauthor{\bsnm{{Costa}}, \binits{F.}},
\bauthor{\bsnm{{Brukner}}, \binits{{\v{C}}.}}:
\batitle{{Universal decoherence due to gravitational time dilation}}.
\bjtitle{Nature Physics}
\bvolume{11}(\bissue{8}),
\bfpage{668}--\blpage{672}
(\byear{2015})
{\href{https://arxiv.org/abs/1311.1095}{{arXiv:1311.1095}}}
{[quant-ph]}.
\doiurl{10.1038/nphys3366}
\end{barticle}
\endbibitem

\bibitem{R2}
\begin{barticle}
\bauthor{\bsnm{{Bassi}}, \binits{A.}},
\bauthor{\bsnm{{Gro{\ss}ardt}}, \binits{A.}},
\bauthor{\bsnm{{Ulbricht}}, \binits{H.}}:
\batitle{{Gravitational decoherence}}.
\bjtitle{Classical and Quantum Gravity}
\bvolume{34}(\bissue{19}),
\bfpage{193002}
(\byear{2017})
{\href{https://arxiv.org/abs/1706.05677}{{arXiv:1706.05677}}}
{[quant-ph]}.
\doiurl{10.1088/1361-6382/aa864f}
\end{barticle}
\endbibitem

\bibitem{diosi2015}
\begin{botherref}
\oauthor{\bsnm{Di\'{o}si}, \binits{L.}}:
Testing spontaneous wave-function collapse models on classical mechanical oscillators
\textbf{114}(5),
050403.
\doiurl{10.1103/PhysRevLett.114.050403}.
Accessed 2023-03-02
\end{botherref}
\endbibitem

\bibitem{vinante2016}
\begin{botherref}
\oauthor{\bsnm{Vinante}, \binits{A.}},
\oauthor{\bsnm{Bahrami}, \binits{M.}},
\oauthor{\bsnm{Bassi}, \binits{A.}},
\oauthor{\bsnm{Usenko}, \binits{O.}},
\oauthor{\bsnm{Wijts}, \binits{G.}},
\oauthor{\bsnm{Oosterkamp}, \binits{T.â.}}:
Upper bounds on spontaneous wave-function collapse models using millikelvin-cooled nanocantilevers
\textbf{116}(9),
090402.
\doiurl{10.1103/PhysRevLett.116.090402}.
Accessed 2023-03-02
\end{botherref}
\endbibitem

\bibitem{bassi2013}
\begin{botherref}
\oauthor{\bsnm{Bassi}, \binits{A.}},
\oauthor{\bsnm{Lochan}, \binits{K.}},
\oauthor{\bsnm{Satin}, \binits{S.}},
\oauthor{\bsnm{Singh}, \binits{T.P.}},
\oauthor{\bsnm{Ulbricht}, \binits{H.}}:
Models of wave-function collapse, underlying theories, and experimental tests
\textbf{85}(2),
471--527.
\doiurl{10.1103/RevModPhys.85.471}.
Accessed 2023-03-02
\end{botherref}
\endbibitem

\bibitem{ghirardi1990}
\begin{botherref}
\oauthor{\bsnm{Ghirardi}, \binits{G.C.}},
\oauthor{\bsnm{Pearle}, \binits{P.}},
\oauthor{\bsnm{Rimini}, \binits{A.}}:
Markov processes in hilbert space and continuous spontaneous localization of systems of identical particles
\textbf{42}(1),
78--89.
\doiurl{10.1103/PhysRevA.42.78}.
Accessed 2023-03-02
\end{botherref}
\endbibitem

\bibitem{diosi1987}
\begin{botherref}
\oauthor{\bsnm{Di\'{o}si}, \binits{L.}}:
A universal master equation for the gravitational violation of quantum mechanics
\textbf{120}(8),
377--381.
\doiurl{10.1016/0375-9601(87)90681-5}.
Accessed 2023-03-02
\end{botherref}
\endbibitem

\bibitem{penrose1996}
\begin{botherref}
\oauthor{\bsnm{Penrose}, \binits{R.}}:
On gravity's role in quantum state reduction
\textbf{28}(5),
581--600.
\doiurl{10.1007/BF02105068}.
Accessed 2023-03-02
\end{botherref}
\endbibitem

\bibitem{oosterkamp2013}
\begin{botherref}
\oauthor{\bsnm{Oosterkamp}, \binits{T.H.}},
\oauthor{\bsnm{Zaanen}, \binits{J.}}:
A clock containing a massive object in a superposition of states; what makes Penrosian wavefunction collapse tick?
{arXiv}.
\url{http://arxiv.org/abs/1401.0176}
Accessed 2023-03-02
\end{botherref}
\endbibitem

\bibitem{Romero2021}
\begin{barticle}
\bauthor{\bsnm{Gonzalez-Ballestero}, \binits{C.}},
\bauthor{\bsnm{Aspelmeyer}, \binits{M.}},
\bauthor{\bsnm{Novotny}, \binits{L.}},
\bauthor{\bsnm{Quidant}, \binits{R.}},
\bauthor{\bsnm{Romero-Isart}, \binits{O.}}:
\batitle{Levitodynamics: Levitation and control of microscopic objects in vacuum}.
\bjtitle{Science}
\bvolume{374}(\bissue{6564}),
\bfpage{3027}
(\byear{2021})
{\href{https://arxiv.org/abs/https://www.science.org/doi/pdf/10.1126/science.abg3027}{{https://www.science.org/doi/pdf/10.1126/science.abg3027}}}.
\doiurl{10.1126/science.abg3027}
\end{barticle}
\endbibitem

\bibitem{leggett2002}
\begin{botherref}
\oauthor{\bsnm{Leggett}, \binits{A.J.}}:
Testing the limits of quantum mechanics: motivation, state of play, prospects
\textbf{14}(15),
415--451.
\doiurl{10.1088/0953-8984/14/15/201}.
Accessed 2023-03-02
\end{botherref}
\endbibitem

\bibitem{arndt2014}
\begin{botherref}
\oauthor{\bsnm{Arndt}, \binits{M.}},
\oauthor{\bsnm{Hornberger}, \binits{K.}}:
Testing the limits of quantum mechanical superpositions
\textbf{10}(4),
271--277.
\doiurl{10.1038/nphys2863}.
Accessed 2023-03-02
\end{botherref}
\endbibitem

\bibitem{Waarde2016}
\begin{botherref}
\oauthor{\bparticle{van} \bsnm{Waarde}, \binits{B.}}:
The Lead Zeppelin: a Force Sensor Without a Handle.
Casimir {PhD} Series,
vol. 2016-28.
Leiden Institute of Physics ({LION}) , Science , Leiden University.
{OCLC}: 966359873
\end{botherref}
\endbibitem

\bibitem{vinante2020}
\begin{botherref}
\oauthor{\bsnm{Vinante}, \binits{A.}},
\oauthor{\bsnm{Falferi}, \binits{P.}},
\oauthor{\bsnm{Gasbarri}, \binits{G.}},
\oauthor{\bsnm{Setter}, \binits{A.}},
\oauthor{\bsnm{Timberlake}, \binits{C.}},
\oauthor{\bsnm{Ulbricht}, \binits{H.}}:
Ultralow mechanical damping with meissner-levitated ferromagnetic microparticles
\textbf{13}(6),
064027
{\href{https://arxiv.org/abs/1912.12252 [cond-mat, physics:physics, physics:quant-ph]}{{1912.12252 [cond-mat, physics:physics, physics:quant-ph]}}}.
\doiurl{10.1103/PhysRevApplied.13.064027}.
Accessed 2023-03-02
\end{botherref}
\endbibitem

\bibitem{aspelmeyer2021}
\begin{botherref}
\oauthor{\bsnm{Westphal}, \binits{T.}},
\oauthor{\bsnm{Hepach}, \binits{H.}},
\oauthor{\bsnm{Pfaff}, \binits{J.}},
\oauthor{\bsnm{Aspelmeyer}, \binits{M.}}:
Measurement of gravitational coupling between millimetre-sized masses
\textbf{591}(7849),
225--228.
\doiurl{10.1038/s41586-021-03250-7}.
Accessed 2023-03-02
\end{botherref}
\endbibitem

\bibitem{brack2022}
\begin{botherref}
\oauthor{\bsnm{Brack}, \binits{T.}},
\oauthor{\bsnm{Zybach}, \binits{B.}},
\oauthor{\bsnm{Balabdaoui}, \binits{F.}},
\oauthor{\bsnm{Kaufmann}, \binits{S.}},
\oauthor{\bsnm{Palmegiano}, \binits{F.}},
\oauthor{\bsnm{Tomasina}, \binits{J.-C.}},
\oauthor{\bsnm{Blunier}, \binits{S.}},
\oauthor{\bsnm{Scheiwiller}, \binits{D.}},
\oauthor{\bsnm{Fankhauser}, \binits{J.}},
\oauthor{\bsnm{Dual}, \binits{J.}}:
Dynamic measurement of gravitational coupling between resonating beams in the hertz regime
\textbf{18}(8),
952--957.
\doiurl{10.1038/s41567-022-01642-8}.
Accessed 2023-03-02
\end{botherref}
\endbibitem

\bibitem{wit2019}
\begin{botherref}
\oauthor{\bparticle{de} \bsnm{Wit}, \binits{M.}},
\oauthor{\bsnm{Welker}, \binits{G.}},
\oauthor{\bsnm{Heeck}, \binits{K.}},
\oauthor{\bsnm{Buters}, \binits{F.M.}},
\oauthor{\bsnm{Eerkens}, \binits{H.J.}},
\oauthor{\bsnm{Koning}, \binits{G.}},
\oauthor{\bparticle{van~der} \bsnm{Meer}, \binits{H.}},
\oauthor{\bsnm{Bouwmeester}, \binits{D.}},
\oauthor{\bsnm{Oosterkamp}, \binits{T.H.}}:
Vibration isolation with high thermal conductance for a cryogen-free dilution refrigerator
\textbf{90}(1),
015112.
\doiurl{10.1063/1.5066618}.
Accessed 2023-03-02
\end{botherref}
\endbibitem

\bibitem{Blakemore2021}
\begin{barticle}
\bauthor{\bsnm{Blakemore}, \binits{C.P.}},
\bauthor{\bsnm{Fieguth}, \binits{A.}},
\bauthor{\bsnm{Kawasaki}, \binits{A.}},
\bauthor{\bsnm{Priel}, \binits{N.}},
\bauthor{\bsnm{Martin}, \binits{D.}},
\bauthor{\bsnm{Rider}, \binits{A.D.}},
\bauthor{\bsnm{Wang}, \binits{Q.}},
\bauthor{\bsnm{Gratta}, \binits{G.}}:
\batitle{Search for non-newtonian interactions at micrometer scale with a levitated test mass}.
\bjtitle{Phys. Rev. D}
\bvolume{104},
\bfpage{061101}
(\byear{2021}).
\doiurl{10.1103/PhysRevD.104.L061101}
\end{barticle}
\endbibitem

\bibitem{smullin2005}
\begin{barticle}
\bauthor{\bsnm{Smullin}, \binits{S.}},
\bauthor{\bsnm{Geraci}, \binits{A.}},
\bauthor{\bsnm{Weld}, \binits{D.}},
\bauthor{\bsnm{Chiaverini}, \binits{J.}},
\bauthor{\bsnm{Holmes}, \binits{S.}},
\bauthor{\bsnm{Kapitulnik}, \binits{A.}}:
\batitle{Constraints on yukawa-type deviations from newtonian gravity at 20 microns}.
\bjtitle{Physical Review D}
\bvolume{72}(\bissue{12}),
\bfpage{122001}
(\byear{2005})
\end{barticle}
\endbibitem

\bibitem{milgrom1983}
\begin{barticle}
\bauthor{\bsnm{Milgrom}, \binits{M.}}:
\batitle{A modification of the newtonian dynamics as a possible alternative to the hidden mass hypothesis}.
\bjtitle{Astrophysical Journal, Part 1 (ISSN 0004-637X), vol. 270, July 15, 1983, p. 365-370. Research supported by the US-Israel Binational Science Foundation.}
\bvolume{270},
\bfpage{365}--\blpage{370}
(\byear{1983})
\end{barticle}
\endbibitem

\bibitem{bekenstein2004}
\begin{botherref}
\oauthor{\bsnm{Bekenstein}, \binits{J.}}:
Relativistic gravitation theory for the mond paradigm. vol. 70.
Phys. Rev. D,
083509
(2004)
\end{botherref}
\endbibitem

\bibitem{carney2021}
\begin{barticle}
\bauthor{\bsnm{Carney}, \binits{D.}},
\bauthor{\bsnm{Krnjaic}, \binits{G.}},
\bauthor{\bsnm{Moore}, \binits{D.C.}},
\bauthor{\bsnm{Regal}, \binits{C.A.}},
\bauthor{\bsnm{Afek}, \binits{G.}},
\bauthor{\bsnm{Bhave}, \binits{S.}},
\bauthor{\bsnm{Brubaker}, \binits{B.}},
\bauthor{\bsnm{Corbitt}, \binits{T.}},
\bauthor{\bsnm{Cripe}, \binits{J.}},
\bauthor{\bsnm{Crisosto}, \binits{N.}}, \betal:
\batitle{Mechanical quantum sensing in the search for dark matter}.
\bjtitle{Quantum Science and Technology}
\bvolume{6}(\bissue{2}),
\bfpage{024002}
(\byear{2021})
\end{barticle}
\endbibitem

\end{thebibliography}

\renewcommand{\appendixtocname}{Supplementary material}

\begin{appendices}

\setcounter{equation}{0}
\setcounter{figure}{0}
\renewcommand{\theequation}{S\arabic{equation}}
\renewcommand{\thefigure}{S\arabic{figure}}

\section{Further Schematics and Photographs of the Setup}

\label{app:photo}
\begin{figure}[ht]
\centering
\includegraphics[width=1\textwidth]{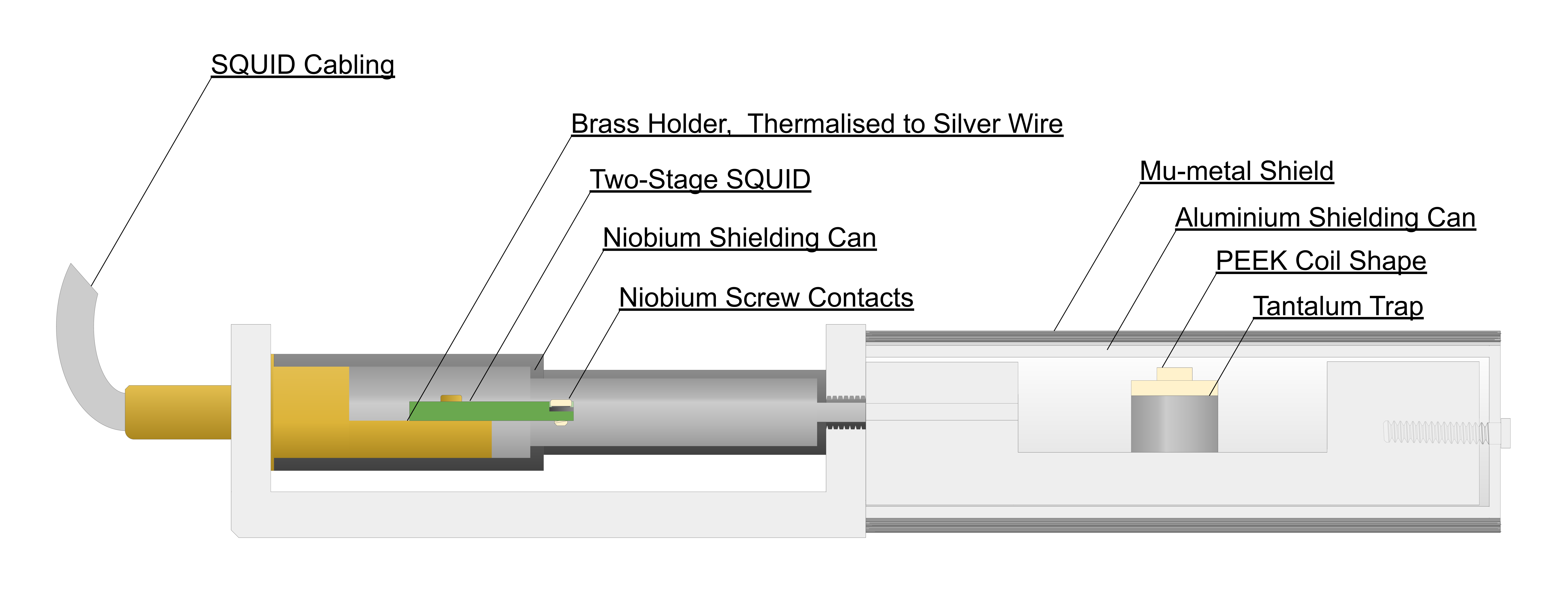}
\caption{\textbf{Close-up of the experiment holder in Fig.~1}. The SQUID detection chip is housed in a niobium can (respectively, Magnicon CAR-1 Two-Stage SQUID and NC-1 Can) that provides shielding from AC magnetic fields through the Meissner effect. The niobium can is screwed into the larger aluminium holder, which similarly provides AC magnetic shielding to the trap through the Meissner effect. The tantalum trap is capped with a PEEK coil-shape, placed offset from the center of the trap, around which the pick-up loop is wound. Additional shielding from DC magnetic fields is provided by several layers of mu-metal foil wrapped around the aluminium holder. This shielding was added under the assumption that stray magnetic fields influence the position of the magnetic particle within the trap and otherwise would get `frozen-in' to the superconductors as they cool down.}\label{fig_supp:holder}
\end{figure}

Since the gravitational coupling is small, great care was taken to shield the experiment from magnetic and electric forces. The heat- and vacuum shields of the cryostat are made of several millimeters of gold plated copper, providing a high level of shielding with respect to electrical forces. Several layers of aluminium foil were wound around the heat shields that achieve temperatures below the critical temperature of aluminium (namely, the \SI{1}{K} shield, and the \SI{50}{mK} shield). This was done in an effort to provide some basic shielding from stray magnetic forces. Further magnetic shielding was incorporated in the holder of the experiment, as shown in figure~\ref{fig_supp:holder}.

\begin{figure}[ht]
\centering
\includegraphics[width=1\textwidth]{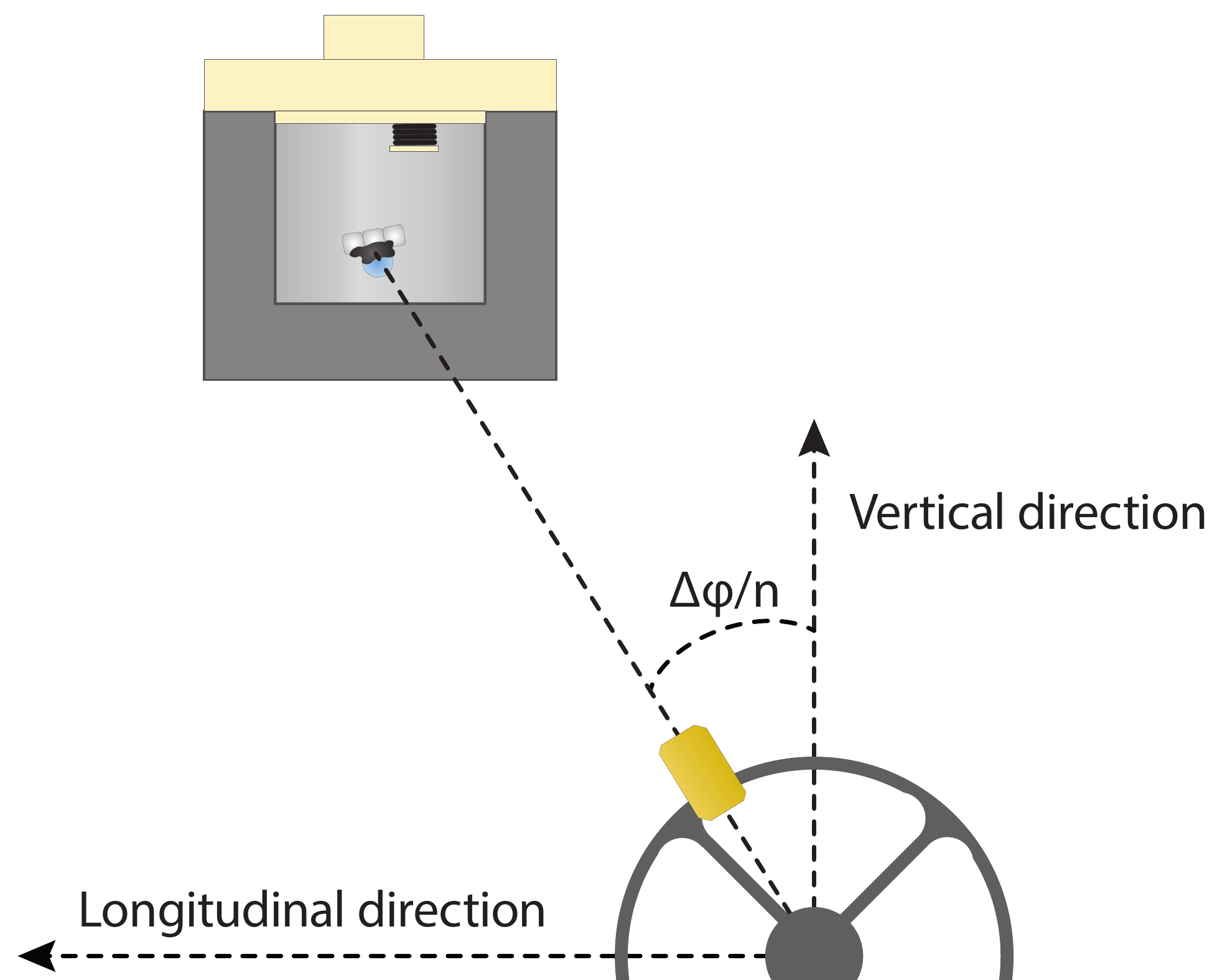}
\caption{\textbf{Schematic depicting the phase angle of the wheel}, as used to define the phase of maximal response in figure 3. Here, $n$ refers to the amount of masses on the wheel, in this case three. This factor follows from the detection not distinguishing between the different masses.}\label{app:phase}
\end{figure}

In figure~\ref{app:phase} we show a schematic to further elucidate the axial system used in the gravitational measurements, showing the longitudinal direction, the vertical direction and the wheel phase over which the wheel was displaced. The factor $n$ follows from the detection by the laser-photodiode combination not distinguishing between masses.

In figure~\ref{fig_supp:photos} are some photographs of critical elements of the experiment. Namely, (part of) the vibration isolation, the coil used to detect the motion of the particle, the transformer used to calibrate the coupling, the tantalum trap, and the wheel used to source the gravitational signal.

\begin{figure}[ht]
\centering
\includegraphics[width=1\textwidth]{Appenidx/supplement_photos.pdf}
\caption{\textbf{Photos of the the experimental apparatus of Fig.1}.\\ \textbf{\emph{A}}: The mass-spring system and the silver wire used for thermalisation of the masses and the experiment, with at the bottom the holder on a small triangular platform to attach the springs. \textbf{\emph{B}}: The coil after the first two layers of five loops were wound around it, with the rest of the cap above. The final pick-up loop consisted of four layers. \textbf{\emph{C}}: A close-up of the calibration transformer, before aluminium shielding was added around it. \textbf{\emph{D}}: The tantalum trap, with the elliptical shape and milling marks visible. \textbf{\emph{E}}: The `mass-wheel' as used in the experiments to show gravitational coupling. The three brass masses are placed in an equilateral triangle. Not shown is the laser-photodiode combination used to read out the frequency of the masses. The wheel is surrounded with one centimeter thick steel plates and is attached to a bridge made out of MK-profiles, which is used to control the elevation and positioning of the wheel in the lab relative to the trap inside the cryostat.}\label{fig_supp:photos}
\end{figure}

\clearpage

\section{Determination of decay time, damping factor, and the quality factor of the mechanical modes}
\label{app:qfactortransfer}
The exponential decay time (\texttau) of each mode was determined through ringdown measurements at high amplitude. At low vibrational amplitude, the measured \texttau\ increases, as is expected from non-linearities in the trapping potential. High amplitude measurements were taken to provide a lower bound on \texttau\ with a high signal-to-noise ratio. We focus on the \SI{27}{Hz} mode, as this is the mode we used to detect our gravitational signal. As discussed in section 2, this mode corresponds most closely to the theoretically expected z-mode frequency. Furthermore, it showed the highest degree of vibrational isolation, which is also as expected given our implemented vibration isolation functions predominantly in z. From the response in phase and amplitude shown in section 2 this is further verified, matching closely to the expected scaling of the z-mode gravitational coupling (both amplitude and phase) under translation of the gravitational source.

\begin{figure}[ht]%
\centering
\includegraphics[width=\textwidth]{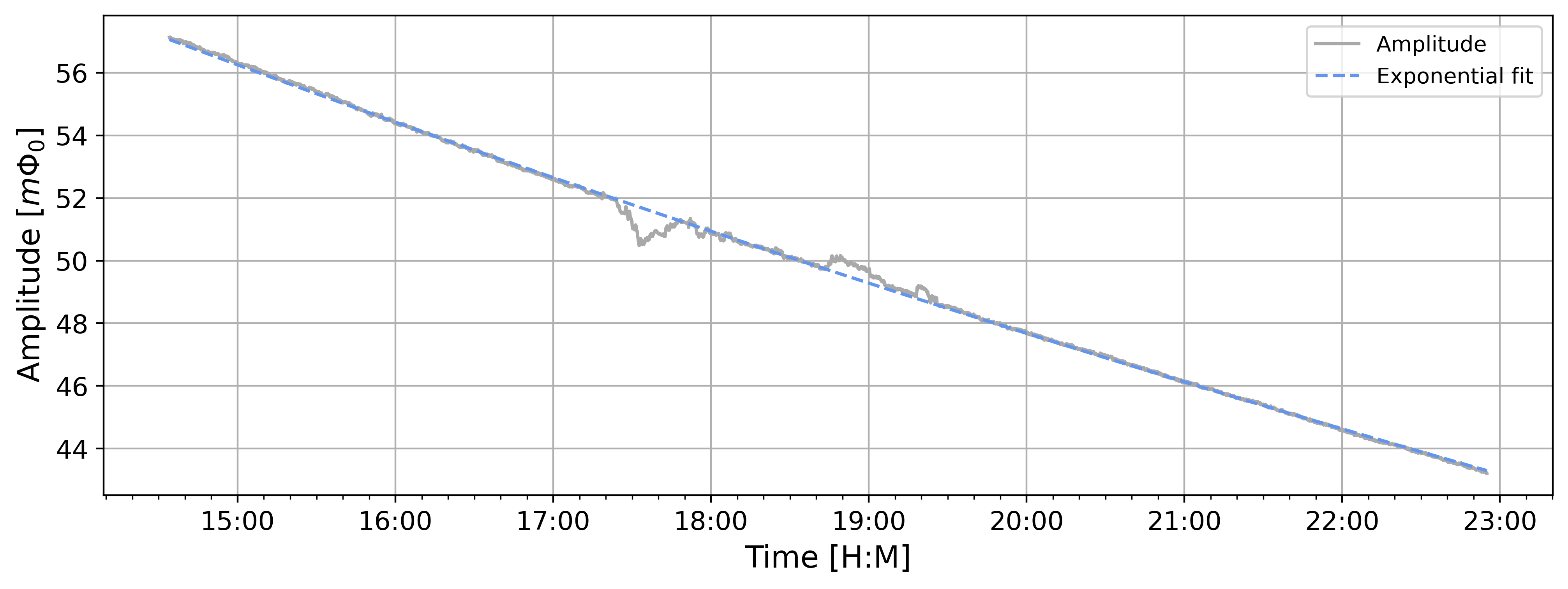}
\caption{\textbf{A typical ringdown of the \SI{27}{Hz} mode}, as performed after a magnetic drive through the calibration transformer. The decay time (\texttau) is extracted from an exponential fit.}\label{fig_supp:ringdown}
\end{figure}

The high amplitude \texttau\ provides a lower bound to the low amplitude \texttau\ of the system. We observe a significant difference in \texttau\ for high and low amplitude, which is explained by a duffing non-linearity in the equations of motion of the resonator. From ringdown measurements, we obtain a lower bound to the decay time of $\tau = \SI{1.09e5}{s}$. The error on the fit of this value is \SI{14.7}{seconds}. At the moment, our understanding of this decay time of individual modes is that it depends strongly on the amplitudes in other modes due to a coupling of the non-linearities. This understanding is limited currently, and a further understanding would require further measurements. To minimise the effect of this false sense of precision (versus accuracy) in the current publication, we will instead truncate the values at three decimals.
With the quality factor (Q) of the resonator defined as

    \begin{equation}
        Q = \pi f \tau 
    \end{equation}

and a frequency of the mode $f = \SI{26.7}{Hz}$, we obtain a quality factor of $Q = \SI{9.06e6}{}$. From \texttau\ we can also determine the damping coefficient of the resonator, which is defined as 

    \begin{equation}
        \gamma = \frac{2}{\tau}
    \end{equation}

which gives a damping in the \SI{27}{Hz} mode of $\gamma = \SI{1.84e{-5}}{s^{-1}}$, or a linewidth of $\gamma/2\pi = \SI{2.92}{\mu Hz}$.

\begin{table}[t]
    \large
    \centering
\begin{tabular}{|c||c|c|} 
 \hline
frequency [Hz] & tau [s] & Q factor\\ [0.5ex] 
 \hline
 15.9 & 3.65$\cdot10^{4}$  & $1.82 \cdot 10^{6}$  \\ 
 \hline
 26.7 & 1.09$\cdot10^{5}$ & $9.13 \cdot 10^{6}$  \\
 \hline
 40.6 & 1.43$\cdot10^{4}$ & $1.82 \cdot 10^{6}$  \\
 \hline
 55.1 & 3.37$\cdot10^{4}$ & $5.84 \cdot 10^{6}$ \\
 \hline
 129 & 0.214$\cdot10^{4}$ & $8.70 \cdot 10^{5}$ \\
 \hline
 147 & 0.152$\cdot10^{4}$ & $6.98 \cdot 10^{5}$ \\
 \hline
\end{tabular}
\caption*{\textbf{Table S1: Tabulated values of the resonator mode parameters.} Again, these values are truncated at at three decimals, since the fit significance of these values is much higher than the actual stability of these values with respect to mode amplitude, as touched upon in the texts. The spring stiffness of the \SI{26.7}{Hz} mode was calculated to be \SI{12}{mN/m}.}\label{tab_supp:tauQ}
\end{table}

\clearpage

\section{Analysis and Calibration of the Single-Stage Particle Readout Circuit and Energy Coupling}
\label{app:calibration}
The motion of the particle, and the resulting force noise of the particle modes, can be calibrated from flux to RMS motion by injecting a magnetic drive through the calibration transformer. The amount of magnetic flux injected is then directly quantised by the SQUID. By measuring the flux induced by the particle response and using the inductance in the circuit and the spring stiffnes, we find a calibration of magnetic flux to motion. 

\begin{figure}[ht]%
\centering
\includegraphics[width=0.5\textwidth]{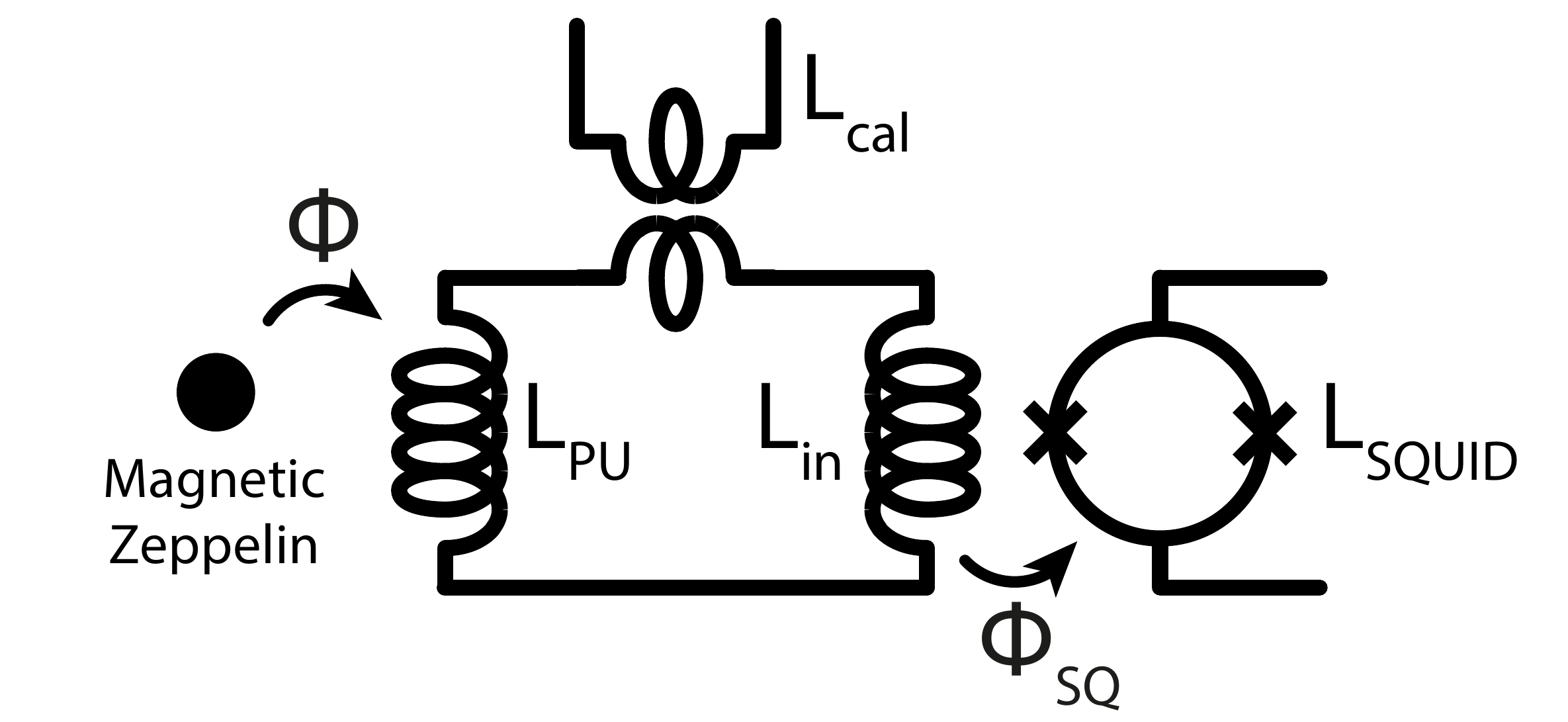}
\caption{\textbf{Schematic circuit of the single-stage readout circuit for the magnetic particle }. The transformer used to couple in the external flux used to calibrate the circuit is depicted at the top, with the particle and pick-up loop on the left, and the input coil and SQUID on the right. }\label{fig_supp:readout_circuit}
\end{figure}

This calibration can be intuitively derived from the energy coupling, $\beta^2$. The energy coupling is defined as the fraction of the energies in two coupled oscillating systems, i.e. the amount of energy that couples from one system to the other. This energy coupling is similar to the quality factor of a single system, which is defined as the fraction of energy lost per cycle with respect to the total energy in the system, or equivalently, a measure for the damping of the system. Conversely, the quality factor provides a measure for the maximal energy stored in the resonator as a fraction of the input energy when driven at resonance for a time $T \gg \tau$, that is to say: the energy at which the fractional energy loss per cycle is equal to the energy put in each cycle by the resonant drive. 

For our system, with a mechanical resonator in the form of the trapped magnetic particle, and a driving magnetic field coupled in through the calibration transformer and coupled to the particle through the pick-up loop, we get 

\begin{equation}
    \beta^2 \equiv \frac{L_{total}I^2}{kx^2} 
\end{equation}

with $x$ any spatial coordinate. Going to the limit of infinitesimal displacement, and using $k =m\omega^2$ and using $L_{total}I^2 = \Phi^2/L_{total}$, we obtain

\begin{equation}
    \beta^2 = \frac{\left(\frac{d\Phi}{dx}\right)^2}{L_{total}\cdot m\omega^2}
\end{equation}

from which it is evident that the energy coupling can be used to determine the absolute motion of the magnetic particle from the flux measured in the SQUID.

To measure $\beta^2$, we further consider the motion of the resonator as a simple resonantly driven harmonic oscillator, with the \emph{Ansatz} $x(t) = A(t)e^{i\omega t};\ A(t) = \frac{F}{2 m \omega}\cdot t$, from which we obtain a damping force

\begin{equation}
    F_{damping} \left(= 2 m \zeta \omega_0 \frac{dx}{dt} = \gamma_m v \right)= \gamma_m \cdot\omega\cdot x
\end{equation}

The quality factor Q is equivalently defined as $Q = \frac{1}{2\zeta}$, thus $\gamma_m = \frac{m\omega}{Q}$.
In our calibration procedure, we inject a flux through the calibration transformer. This flux then results in a current through the detection circuit. This current is detected as a flux by the SQUID, with known $\frac{dI}{d\phi}$ calibration from the SQUID parameters. We refer to this current as the the crosstalk current $I_{crosstalk}$. This current also leads to a flux through the pick-up loop, which gives rise to a magnetic force on the particle of the form $F_{drive}=\alpha \cdot I_{crosstalk}$. Here, $\alpha$ is a coupling constant of units $\si{[N/A]}$ that contains a geometrical factor determined by both the relative positioning of the particle with respect to the pick-up coil, and the physical sizes of this coil and particle. As the particle is driven, the motion of the particle will in turn induce a current in the detection circuit. This induced current has the form $I_{induced}=\delta\cdot x$ where $\delta$ is a coupling constant with units $\si{[A/m]}$, which is by symmetry effected through the pick-up coil-particle geometry in the same way that $\alpha$ is.\\
Combining these results, and noting that for the steady state solution the driving force must be equal to the damping force

\begin{equation}
    \frac{I_{induced}}{I_{crosstalk}} =
    \frac{\delta\cdot x}{F_{drive}/\alpha} =
    \frac{\delta \cdot \alpha\cdot x}{\gamma_m \cdot \omega \cdot x} =
    Q\cdot \frac{\alpha\cdot\delta}{m\omega^2} = 
    Q\cdot \frac{\alpha\cdot\delta}{k}
    \label{eq:I/I}
\end{equation}

Since some of our decay times are rather long, in our experiments we work with a $Q_{eff} = \pi fT$, where $T$ is the time during which the drive is applied. Whilst this is not the steady state, this discrepancy is fully absorbed in our definition of $Q_{eff}$:

\begin{equation}
    \frac{I_{induced}}{I_{crosstalk}} =
    \frac{\delta\cdot x}{F_{drive}/\alpha} =
    \frac{\delta \cdot \alpha\cdot x}{\gamma_{eff} \cdot \omega \cdot x} =
    Q_{eff}\cdot \frac{\alpha\cdot\delta}{m\omega^2} = 
    Q_{eff}\cdot \frac{\alpha\cdot\delta}{k}
\end{equation}

From this derivation, we have found that the fraction of the currents in our detection circuit contains all the coupling constants in our system. In fact, looking at the units, we find

\begin{equation}
    \frac{\alpha\cdot\delta}{k} = 
    \frac{\si{\left[\frac{N}{A}\right]}\cdot\si{\left[\frac{A}{m}\right]}}{\si{\left[\frac{N}{m}\right]}}
\end{equation}

Furthermore, for small displacements in $x$, the flux change through the pick-up loop can be treated as linear. The resultant current in the detection circuit must then be $I_{induced} = \frac{d\Phi}{dx}\cdot x \cdot \frac{1}{L_{total}}$ or $\delta = \frac{d\Phi}{dx}\cdot \frac{1}{L_{total}} \hat{=} \si{\left[\frac{A}{m}\right]}$.

As noted before, $\delta$ contains the same geometric scaling as $\alpha$. From the symmetry, we recognise that $\alpha = \frac{d\Phi}{dx}$ ($\left[\SI{}{Wb/m}\right]= \left[\SI{}{N/A}\right]$). Combining these results, we find

\begin{equation}
    \frac{\alpha\cdot\delta}{k} = \frac{\left(\frac{d\Phi}{dx}\right)^2}{L_{total}\cdot m\omega^2} = \beta^2
\end{equation}

This final results gives us a measurable result for $\beta^2$, or, conversely, the proportionality constant $\frac{d\Phi}{dx}$ that equates our measured flux to an absolute motion. The calibration measure is then found from 

\begin{equation} \label{eq:dphi_dx_beta}
    \frac{d\Phi}{dx} = \sqrt{L_{total}\cdot m\omega^2\cdot \beta^2} = \sqrt{L_{total}\cdot m\omega^2\cdot \frac{I_{induced}}{Q\cdot I_{crosstalk}}} 
\end{equation} 

Noting that both $I_{induced}$ and $I_{crosstalk}$ are converted equally by the SQUID, the measured voltage can be inserted for the current through the detection circuit, which provides our sensitivity as 

\begin{equation}
    \frac{d\Phi}{dx} = \sqrt{L_{total}\cdot m\omega^2\cdot \frac{\Delta V_{drive}}{Q\cdot V_{crosstalk}}}
\end{equation}

Here $\Phi$ is the flux through the pick-up loop. Since we detect this using the SQUID, we need to convert this value to the flux through the SQUID, which is done by dividing by the total inductance of the circuit and multiplying by the mutual inductance of the SQUID input loop and the SQUID loop. As a final step, this value can be related directly to the voltage measures made by taking the SQUID gain value as measured.  

\begin{equation}
    \frac{dV}{dx} = 
    \frac{dV}{d\Phi_{SQ}}\cdot\frac{d\Phi_{SQ}}{dI}\cdot\frac{dI}{d\Phi}\cdot\frac{d\Phi}{dx} =
    \frac{dV}{d\Phi_{SQ}} \cdot M_{in,SQ} \cdot \frac{1}{L_{total}}\cdot \frac{d\Phi}{dx}
\end{equation}

The values for the inductances in $L_{total} = L_{PU}+L_{TP}+L_{input}+L_{calibration}$ are: $L_{PU} = \SI{2.9e{-7}}{H}$ the inductance of the pick-up loop, $L_{TP} = \SI{1e{-7}}{H}$ the inductance of the \SI{10}{cm} superconducting twisted-pair connecting the SQUID input to the pick-up loop, $L_{input} = \SI{4e{-7}}{H}$ the inductance of the input coil of the two-stage SQUID device, and $L_{calibration} = \SI{2e{-9}}{H}$ the inductance of the small calibration transformer coil. For the mutual inductance, the SQUID is calibrated to a value of $M_{in,SQ}^{-1} = \SI{0.5}{\mu A/\Phi_0}$. The SQUID voltage gain is measured as $dV/d\Phi_0 = \SI{0.43}{V/\Phi_0}$. These values, combined with measures of these ring-up's, gives us a value of $\frac{dV}{dx} = \SI{0.16}{{\frac{V}{\mu m}}}$ for the $\SI{27}{Hz}$ mode, with a relative error of 7\%, which is dominated by the error in the voltage calibration measurement. 

\newpage

\section{Comparison to basic Optomechanical Quantities}
Given the vast previous work in the field of optomechanics, we aim to provide a rough outline of how the work presented in this paper compares to quantities in optomechanical experiments, in this appendix. For a more complete treatise of optomechanics and specifically a comparison to magneto-mechanics, we refer to \cite{Romero2021}.

We note that the resonance frequency of our levitated particle corresponds to the frequency of the mechanical oscillator in an optomechanics experiment. At present, our experiment lacks the equivalent of the optical cavity and the light sent to the cavity. It could be added to the experiment, by changing the readout of our experiment. Currently, we use a SQUID based read-out, that relies on DC signals. We could read out the SQUID at higher frequencies (using a GHz cavity). In that case the inductance of the SQUID, which depends on the flux in the SQUID, would provide frequency tuning to a LC resonator, with typical resonance frequencies of 2-20 GHz. 

We adopted $\omega$ to be the frequency of the mechanical resonator, this in contrast to optomechanics, in which $\Omega$ is nominally adopted for the mechanical frequency, with $\omega$ reserved for the optical frequency. In the rest of this appendix, we will refer to the optical frequency as used in optomechanics as $\omega_{opt}$ to avoid confusion.
In the context of optomechanics, the optomechanical coupling, g, is given by the infinitesimal change in wavelength as a function of the change in position of the mechanical oscillator:

    \begin{equation}
        g_{opt} = \frac{d\omega_{opt}}{dx}
    \end{equation}

This is an important quantity that provides a measure on how well the mechanical and optical systems are coupled. Often the value is quoted in terms of the zero-point-motion ($x_{zpm}$) of the resonator, as $g_0$.
In the case of magnetic levitation, a translation from  $\frac{d\Phi}{dx}$ to $\frac{d\omega_{opt}}{dx}$ can be made through the realization that the inductance of the SQUID $L_J=\frac{\Phi_0}{2 \pi I_c(\Phi)}$ depends on the flux through the SQUID. This change in inductance changes the resonance frequency of the LC resonator, $\omega_{LC}$, that could be used to readout our SQUID in the future. In the same way that the resonance frequency of an optical cavity depends on the position of a membrane, we also see that the resonance frequency of the LC circuit that the SQUID is part of, depends on the position of our levitated magnetic particle. 

Combining this with equation~\ref{eq:dphi_dx_beta}, we find:

\begin{equation}
    g_{mag} = \frac{d\omega_{LC}}{d\Phi} \frac{d\Phi}{dx} = \frac{d\omega_{LC}}{d\Phi} \sqrt{L_{total}\cdot m\omega^2\cdot \beta^2}
\end{equation}

Evidently, this is the equivalent to $g_{opt}$ for our SQUID detected system. Implicit in the term $L_{total}$ is the coupling loss from parasitic inductances. 
Since the detection is not resonant, like it would be in an optical cavity system, this remains a negligible loss. 
Expressing $g$ in terms of the resonator zero-point-motion can be done similar to how it is often derived in optomechanical systems, from equating the energy of the bosonic system at zero occupation to the kinetic energy at zero point motion:

\begin{equation}
    x_{zpm} = \sqrt{\frac{\hbar}{2 \cdot m_{eff} \cdot \omega_m}}
\end{equation}

Where $m_{eff}$ is the effective mass of the mechanical oscillator and $\omega_m$ is the frequency of the mechanical oscillator. In our case the $m_{eff}$ is just the mass of the levitated particle and $\omega_m$ is the eigenfrequency of the measurement. Filling in the mass and frequency used in the gravity measurements, a zero point motion of $x_{zpm} = \SI{0.86}{fm}$ can be calculated. Combining this with our measured sensitivity, we find $\frac{d\Phi}{dx} \cdot x_{zpm}= \SI{63}{n\Phi_0}$ as a zero point flux for our system. For a slope, $\frac{d\omega_{LC}}{d\Phi}$,  of a $\lambda/4$ resonator terminated with a SQUID of $\SI{1}{GHz/\Phi_0}$, this results in a $g_{0,mag} = g_{mag} \cdot x_{zpm} = \SI{63}{Hz}$. 
We conclude, however, that it will be a considerable challenge to make the Q factor of such a LC resonator high enough to reach the sideband resolved limit.



\newpage

\section{Correction and Conversion of the Data to Units of \texorpdfstring{$\SI{}{N/\sqrt{Hz}}$}{N/rtHz}}
\label{app:correction_and_conversion}
The data used in our gravitational experiments comes from the photodiode detector which measures the passages of the masses on the wheel, and the SQUID signal. Both these signals are filtered by our lock-in amplifiers. Since our gravity experiments were performed at differing amplitudes in the mode, each time trace had a different slope, as discussed in Supp.~\ref{app:qfactortransfer}. To account for this, we have subtracted an individual exponential decay in the form of $Ae^{t/\tau_{fit}}$ from each.

\begin{figure}[ht]
\centering
\includegraphics[width=\textwidth]{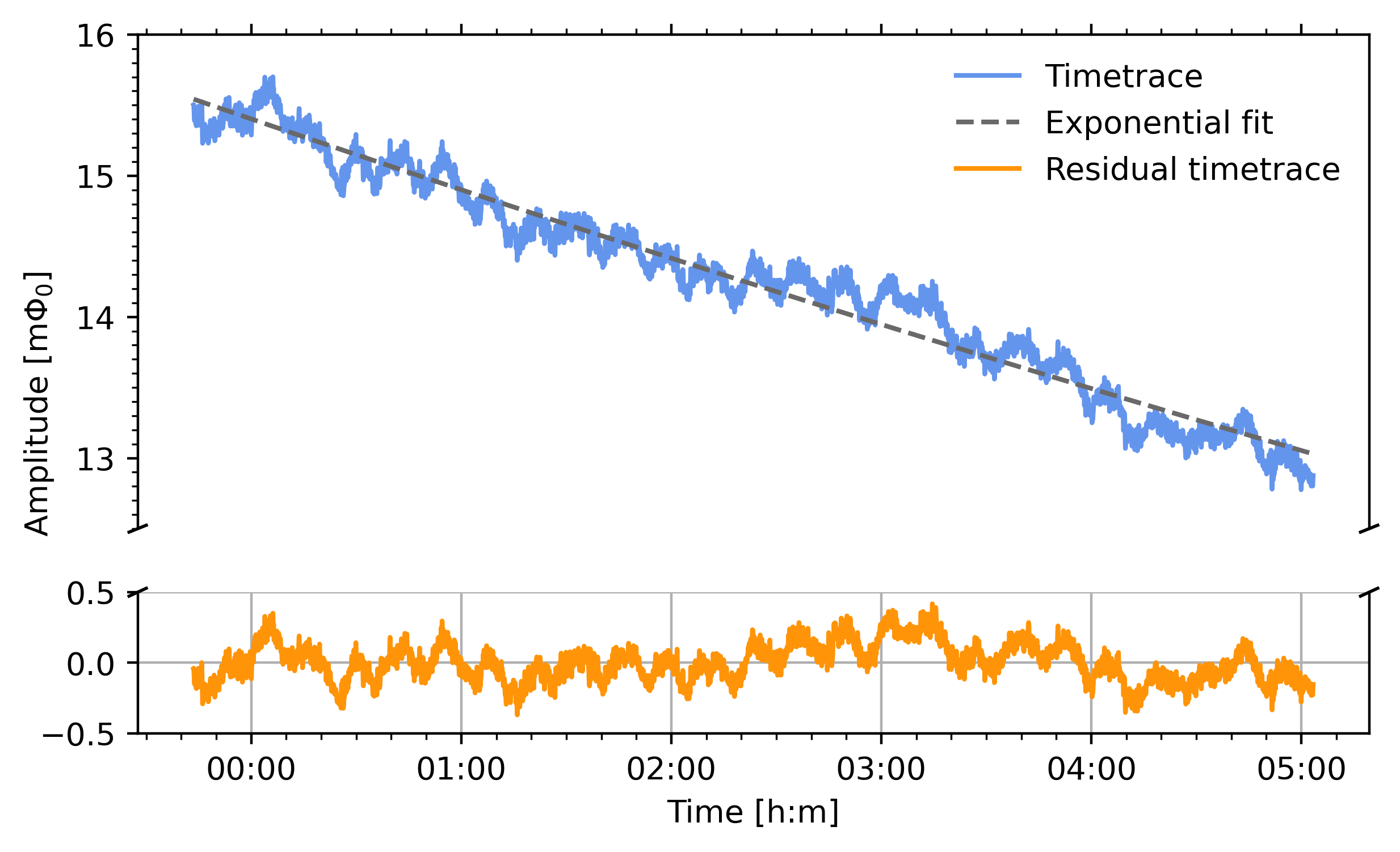}
\caption{\textbf{Ringdown subtraction as performed on the time trace measurement of each wheel position}. Shown here for a separation between particle and driving mass at wheel phase zero of \SI{48.1}{cm} vertical, and laterally displaced by \SI{3.5}{cm}. Residual plotted in the lower half, where we see the fluctuations as result of the drive up and drive down effected by the mass-wheel, which is slightly detuned in frequency with respect to the mode. We drive detuned since we wish to stay away from the non-linear driving and frequency shift of the mode that happens at large amplitude. Furthermore, detuned driving enables us to substract the absolute amplitude.}
\end{figure}

After subtracting the ringdown, we apply a phase factor to the time traces. This phase factor is determined based on the detuning of the resonator frequency with respect to the lock-in center frequency. This change ensures that the central peak of the resonator mode will fall fully in a single bin of the Fourier transform we perform next, this ensures that the transfer function of the resonator mode is fully symmetrical. Before performing the Fourier transform, we also cut out a section of the time trace in which there is an integer amount of phase cycles for the mass-passage signal, which ensures smooth periodic boundary conditions for our FFT.
This results in a spectrum giving us the motion of the particle per root hertz, which can be converted to RMS motion in a specific frequency band by integrating over that frequency bandwidth.

\begin{figure}[ht]
\centering
\includegraphics[width=\textwidth]{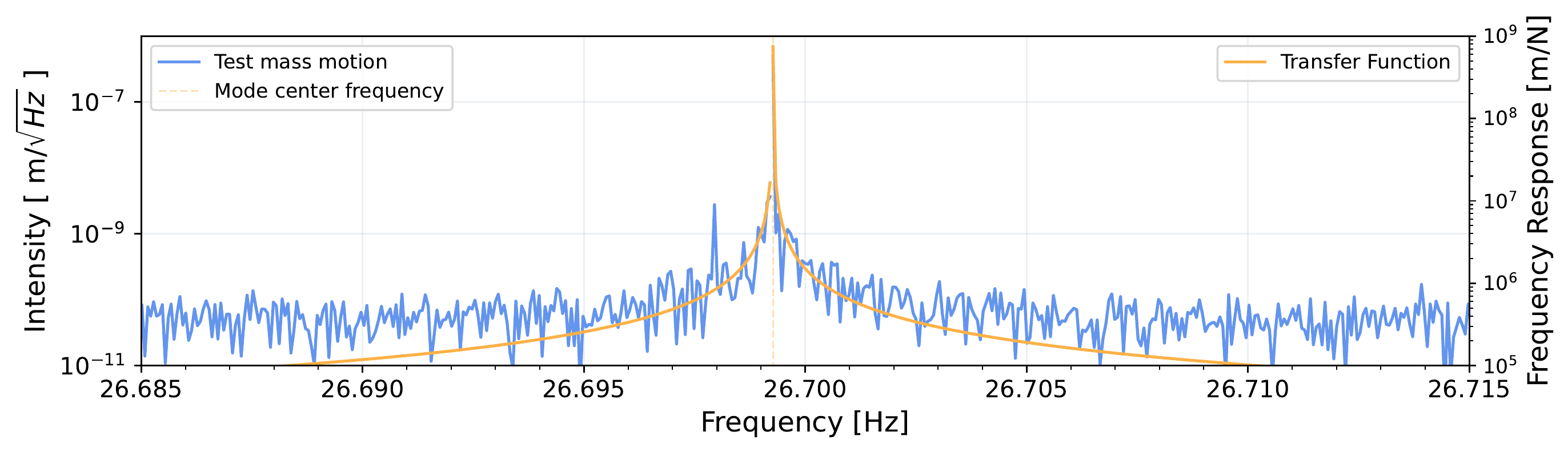}
\caption{\textbf{Spectrum of the time trace converted to motional noise, with the transfer function plotted overlaid}. The orange vertical line indicates the resonance frequency of the mode.}
\end{figure}

By then subtracting the transfer-function of the mechanical mode from the particle signal, and applying our conversion factor to go from $\Phi_0$ to displacement, and using a spring stiffness $k = m\omega^2$ and a mode bandwidth of $df=Q/f$, we arrive at our force noise spectrum in $\SI{}{N/\sqrt{Hz}}$

\begin{figure}[ht]%
\centering
\includegraphics[width=\textwidth]{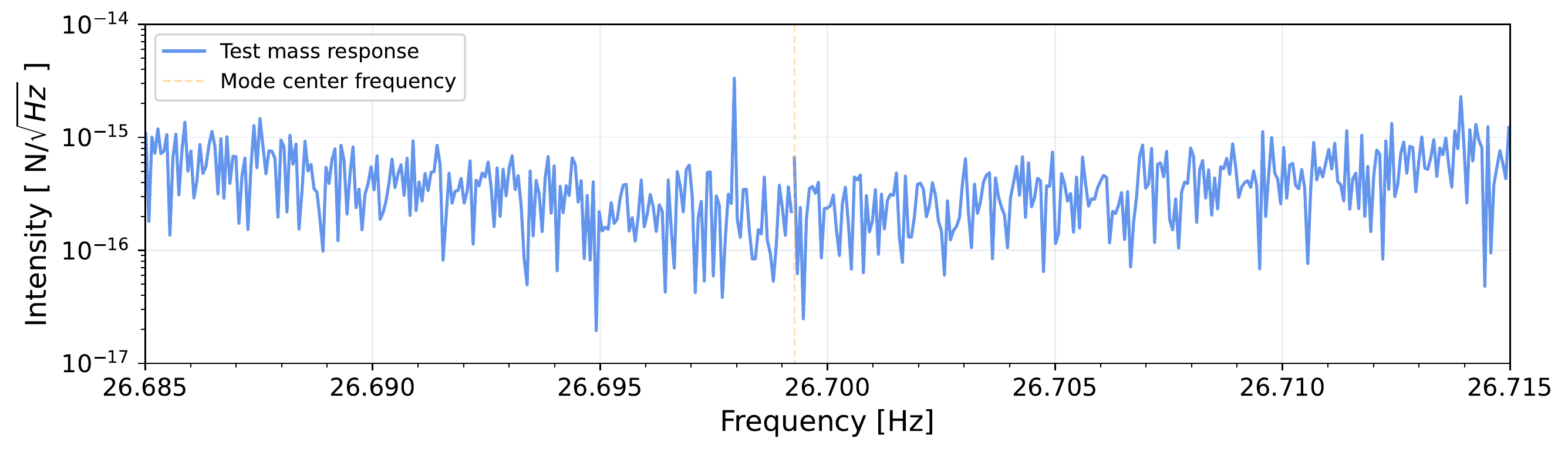}
\caption{\textbf{Force noise spectrum}. Spectrum of the time trace converted to force noise, the final product of this procedure. The orange vertical line indicates the resonance frequency of the mode.}
\end{figure}

The noise floor of this measurement corresponds to a mode temperature of \SI{3}{K}, from $T_{mode} = kx_{RMS}^2/k_B$.

\end{appendices}

\end{document}